\documentclass[prb,twocolumn,showpacs]{revtex4}
\usepackage{bm}
\usepackage{graphicx}
\usepackage{epsfig}

\begin{document}

\title{Spin-dependent resonant tunneling in
symmetrical double-barrier structure}

\author{M.M.~Glazov}
\author{P.S.~Alekseev}
\author{M.A.~Odnoblyudov}
\author{V.M.~Chistyakov}
\author{S.A.~Tarasenko}\email{tarasenko@coherent.ioffe.ru}
\author{I.N.~Yassievich}
\affiliation{A.F.~Ioffe Physico-Technical Institute, RAS, 194021
St.~Petersburg, Russia }

\date{\today}

\begin{abstract}
A theory of resonant spin-dependent tunneling has been developed
for symmetrical double-barrier structures grown of
non-centrosymmetrical semiconductors. The dependence of the
tunneling transparency on the spin orientation and the wave vector
of electrons leads to (i) spin polarization of the transmitted
carriers in an in-plane electric field, (ii) generation of an
in-plane electric current under tunneling of spin-polarized
carriers. These effects originated from spin-orbit
coupling-induced splitting of the resonant level have been
considered for double-barrier tunneling structures.
\end{abstract}


\maketitle
\section{Introduction}
Physics of spin-dependent tunneling phenomena in semiconductor
structures has attracted lately a great deal of attention.
Significant progress has been made in experimental and theoretical
investigation of spin-polarized transport in magnetic tunneling
junctions (for review see~[\onlinecite{Tsymbal,Zutic}]). On the
other hand, it was pointed out recently that electron tunneling
could be spin-dependent even in the case of {\it nonmagnetic}
barriers. It was demonstrated that the transparency of a
semiconductor barrier depends on the spin orientation of carriers
if the system lacks a center of inversion. Two microscopic
mechanisms, Rashba spin-orbit coupling induced by the barrier
asymmetry~\cite{Zakharova,Voskoboynikov,Silva} and the $k^3$
Dresselhaus spin-orbit splitting in non-centrosymmetrical
materials,\cite{Perel,Botha} were shown to be responsible for the
effect of spin-dependent tunneling. Spin-orbit interaction couples
spin states and space motion of conduction electrons that opens a
possibility to orient, manipulate and detect spins by electrical
means. Effect of spin-dependent tunneling in nonmagnetic
semiconductor heterostructures was supposed to be applied for spin
injection~\cite{Voskoboynikov2,Ting,Koga,Shang} and detection of
spin-polarized carriers.\cite{Hall,Tarasenko} Devices based on
spin-dependent tunneling were suggested to be utilized as
components of the spin field-effect transistor.\cite{Hall}

In this paper we present a theory of spin-dependent tunneling
through a double-barrier structure, the structure of resonant a
tunnel diode (RTD), grown of non-centrosymmetrical semiconductors.
Section~\ref{transmission} is devoted to calculation of the
spin-dependent transmission coefficient. The dependence of the
structure transparency on the spin orientation and the wave vector
of electrons can be employed for spin injection and detection: (i)
an electric current flow in the plane of interfaces leads to the
spin polarization of the transmitted carriers, (ii) transmission
of the spin-polarized carriers is accompanied by generation of an
in-plane electric current. These effects are considered in
Sections~\ref{orientation}~and~\ref{detection}, respectively. The
results of the numerical calculations are compared with that
obtained in a simple analytical theory.
\section{Spin-dependent resonant tunneling}\label{transmission}
\begin{figure}[b]
\vspace{0.5cm}
\includegraphics[width=0.45\textwidth]{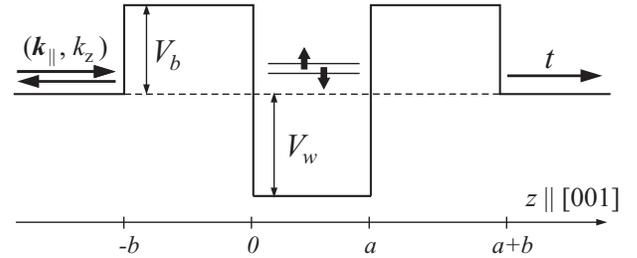}\\
\caption{Electron transmission through (001)-grown RTD structure.
$V_b$ is the height of the barriers, $V_w$ is the depth of the
well, $a$ and $b$ are the well width and barrier thickness,
respectively.}\label{Fig1}
\end{figure}

We consider the transmission of electrons with the initial wave
vector $\bm{k}=(\bm{k}_{\parallel},k_z)$ through a symmetrical
double barrier structure grown along $z \parallel [001]$ direction
(see Fig.\ref{Fig1}).
Here $\bm{k}_{\parallel}$ is the wave
vector in the plane of the interfaces, and $k_z$ is  wave vector
component normal to the barrier and pointing in the direction of
tunnelling. The electron motion in each layer of the structure is
described by the Hamiltonian
\begin{equation}\label{h1}
\hat H = - \frac{\hbar^2}{2m^*} \frac{\partial^2}{\partial z^2} +
\frac{\hbar^2 k_{\parallel}^2}{2m^*} + V(z) + \hat{\mathcal
H}_{D},
\end{equation}
where $m^*$ is the electron effective mass, $V(z)$ is the
heterostructure potential equal to $V_b$ for the barriers and
$-V_w$ for the well, and $\hat{\mathcal H}_{D}$ is the
spin-dependent $k^3$ Dresselhaus term that describes spin-orbit
splitting of the conduction band in zinc-blende-lattice
semiconductors. We assume the RTD structure to be designed so that
the resonant transmission occurs for the incident electrons with
the kinetic energy $\varepsilon$ much smaller than $V_b$ and
$V_w$. Then the Dresselhaus term is simplified to~\cite{Perel}
\begin{equation}\label{H_D1}
\hat{\mathcal H}_{D} = \gamma (\hat{\sigma}_x k_x -
\hat{\sigma}_y k_y) \frac{\partial^2}{\partial z^2}  \:,
\end{equation}
where $\gamma$ is a material constant, $\hat{\sigma}_{x}$ and
$\hat{\sigma}_{y}$ are the Pauli matrices, and the coordinate axes
$x,y,z$ are assumed to be parallel to the cubic crystallographic
axes $[100]$, $[010]$, $[001]$, respectively.

The Dresselhaus term~(\ref{H_D1}) is diagonalized by the spinors
\begin{equation}\label{chi_pm}
\chi_{\pm} = \frac{1}{\sqrt{2}}\left(
\begin{array}{c}
1 \\ \mp \mbox{e}^{- i \varphi}
\end{array}
\right) \:,
\end{equation}
which describe the electron states ``$+$'' and ``$-$'' of the
opposite spin directions. Here $\varphi$ is the polar angle of the
wave vector $\bm{k}$ in the $xy$ plane, $\bm{k}_\parallel =
(k_{\parallel} \cos \varphi \,, \: k_{\parallel} \sin \varphi)$.
The electron spins $\bm{s}_\pm$ corresponding to the eigen-states
``$\pm$'' are given by
\begin{equation}\label{s_dr}
\bm{s}_{\pm}(\bm{k}_\|) = \frac12 \, \chi^{\dag}_{\pm} \hat{ \bm
\sigma} \chi_{\pm} = \frac12 \, (\mp \cos \varphi \,, \: \pm \sin
\varphi \,, \: 0) \:.
\end{equation}

One can note that essentially $\hat{\mathcal H}_{D}$ induces a
spin-dependent correction to the effective mass for electron
motion along $z$. In the basis of the spin eigen-states ``$\pm$''
the effective Hamiltonian~(\ref{h1}) has the simple form
\begin{equation}\label{heff}
H_{\pm} = - \frac{\hbar^2}{2m_{\pm}} \frac{\partial^2}{\partial
z^2} + \frac{\hbar^2 k_{\parallel}^2}{2m^*} + V(z),
\end{equation}
where the effective mass along $z$, modified by spin-orbit
coupling, depends on the in-plane electron wave vector and is
given by
\begin{equation}
m_{\pm} = m^* \left(1 \pm 2\frac{\gamma m^*
k_{\|}}{\hbar^2}\right)^{-1} .
\end{equation}
Solution of the Schr\"odinger equation with the
Hamiltonian~(\ref{heff}) and boundary conditions for the wave
functions $\psi_{\pm}$, which require the continuity of
\[
\psi_{\pm} \hspace{1cm} \mbox{and}    \hspace{1cm}
\frac{1}{m_{\pm}}\frac{
\partial \, \psi_{\pm}}{ \partial \, z}
\]
at the interfaces, allows one to derive coefficients of
transmission, $t_\pm$, and reflection, $r_\pm$, for the electrons
of spin eigen-states ``$+$'' and ``$-$''.

\begin{figure}[t]
\vspace{0.3cm}
\includegraphics[width=0.43\textwidth]{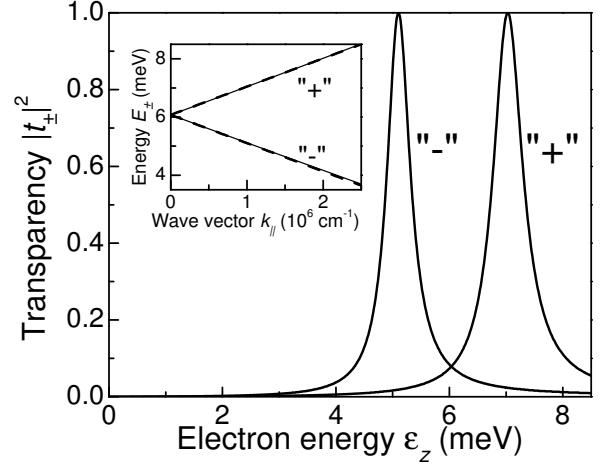}\\ \caption{The
transparency of the double-barrier structure, $|t_\pm|^2$, as a
function of $\varepsilon_z$ at fixed $k_\parallel =
10^6$~cm$^{-1}$. The insert shows the dependence of the spin
splitting of the resonant peak on $k_\parallel$ calculated
numerically (solid curves) and following Eq.(\ref{expansion})
(dashed lines). The used parameters, $\gamma =
76$~eV$\cdot$\AA$^{3}$, $m^*=0.053m_0$, $V_b=230$~meV,
$V_w=200$~meV, $a=30$\AA, and $b=50$\AA, correspond to
Al$_{x}$Ga$_{1-x}$Sb, $x=0.15/0.3/0/0.3/0.15$, RTD
structure.~\cite{non_parabolicity}}\label{Fig2}
\end{figure}

Fig.\ref{Fig2} presents the dependencies of the double-barrier
structure transparency $|t_\pm|^2$ on the incident electron energy
along growth direction, $\varepsilon_z=\hbar^2 k^2_z / 2 m^*$,
calculated numerically for the fixed in-plane wave vector
$k_\parallel$. The spin splitting of the resonant peak is clearly
seen. In calculations both the electron effective mass $m^*$ and
the Dresselhaus constant $\gamma$ are assumed to be the same for
the barrier and well layers.

In the limit of thick barriers the structure transparency
demonstrates sharp peaks and hence can be approximated by Dirac
$\delta$-functions
\begin{equation}\label{t}
\left|t_\pm(\varepsilon_z, k_{\|})\right|^2 \, \approx \, \pi
\Gamma_\pm({ k_\|})\, \delta\left[\varepsilon_z-E_{\pm}(
k_{\|})\right],
\end{equation}
where the prefactors $\Gamma_\pm({ k_\|})$ describe the
transmission efficiencies, and $E_{\pm}( k_{\|})$ stand for the
energies of the resonances. The positions of the resonances
$E_{\pm}( k_{\|})$ correspond to the energies of size-quantization
of an electron in the quantum well of infinitely thick barriers
with the Dresselhaus spin-orbit interaction included. The
transmission efficiencies $\Gamma_\pm({ k_\|})$ are determined, to
the contrary, by the electron lifetimes on the resonant levels in
the double-barrier structure. Considering the spin-orbit
interaction to be small perturbation, the positions of the
resonant levels and their widths can be expanded as
\begin{equation}\label{expansion}
E_{\pm}({ k}_\parallel) = E_0 \pm \alpha k_\parallel \:, \quad
\Gamma_{\pm}({ k}_\parallel) = ( 1 \pm \beta k_\parallel ) \,
\Gamma_0 \:,
\end{equation}
where $E_0$ and $\Gamma_0$ are the level position and the width
when the spin-orbit interaction is neglected,
\begin{equation}
\Gamma_0=8 \sqrt{E_0(V_b-E_0)} \:
\frac{(V_b-E_0)(V_w+E_0)}{V_b(V_b+V_w)} \, \frac{\exp(-2\kappa
b)}{ 1+\kappa a/2} \:,
\end{equation}
$\kappa= \sqrt{2m^*(V_b-E_0)}/\hbar$ is the reciprocal length of
the wave function decay under the barrier, $a$ is the quantum well
width, $b$ is the thickness of the barriers, and $\alpha$ and
$\beta$ are the coefficients which describe the spin splitting of
the level and the width modification of the spin sublevels,
respectively. For the case under study, $E_0 \ll V_b, V_w$, they
are given by
\begin{equation}
\alpha= \frac{2\gamma m^*}{\hbar^2} \frac{V_w}{1 + 2/\kappa a} \:,
\end{equation}
\[
\beta = \alpha \left( \frac{\kappa b}{V_b} + \frac{1}{2E_0}
\right)+ 2\gamma \frac{m^* \kappa b}{\hbar^2} \:.
\]
Note, that the width $\Gamma_0$ and the coefficient $\beta$ are
highly sensitive to the position of the resonant level $E_0$. For
the RTD structure parameters presented in the caption to
Fig.\ref{Fig2}, the coefficients can be estimated as follows:
$\alpha=9.7\cdot10^{-7}$~meV$\cdot$cm,
$\beta=(4.2\cdot10^{-8}+4.8\cdot10^{-7}\cdot{\rm meV}/E_0)$~cm.
\section{Spin orientation of carriers}\label{orientation}
Spin splitting of the resonant peak at non-zero $\bm
k_{\parallel}$ can be employed for injection of spin-polarized
carriers. We assume two parts of bulk semiconductor separated by
the RTD structure, and electrons tunneling through the barrier
from the left to the right side of the structure. In equilibrium
the momentum distribution of the incident electrons is isotropic
in the interface plane and therefore the average spin of the
transmitted carriers vanishes. This isotropy can be broken by
application, for example, of an in-plane electric field, $\bm F$.
Then the carriers tunnel through the structure with non-zero
average wave vector in the plane of interfaces that leads to the
spin polarization of the transmitted
electrons.\cite{Voskoboynikov2,Ting}

The average spin of the transmitted electrons is given by
\begin{equation}\label{polarization}
\bm s = \dot{\bm S} / \dot N \:,
\end{equation}
where $\dot{\bm S}$ and $\dot N$ are the spin and carrier fluxes
through the barrier given in the linear in $\bm F$ regime by
\begin{equation}\label{spin_gen_f}
\dot{ \bm S} = \sum_{\bm k_\|, \, k_z>0} f_1 (\bm k) \left[ \left|
t_{+}(\varepsilon_z, k_{\|}) \right|^2 \bm s_{+} (\bm k_{\|})
\right.
\end{equation}
\vspace{-0.7cm}
\[
\hspace{3cm} +\, \left. \left| t_{-}(\varepsilon_z, k_{\|})
\right|^2 \bm s_{-}(\bm k_{\|}) \right] v_z \:, \nonumber
\]
\[
\dot N = \sum_{\bm k_\|, \, k_z>0}  f_0(\varepsilon)\left[ \left|
t_{+}(\varepsilon_z, k_{\|}) \right|^2 + \left|
t_{-}(\varepsilon_z, k_{\|}) \right|^2\right] v_z\:,
\]
$\bm v = \hbar \bm k / m^*$ is the electron velocity,
$f_0(\varepsilon)$ is the equilibrium distribution function,
$\varepsilon$ is the electron energy, $f_1 (\bm k)$ is the
electric field-induced correction to the distribution function,
\[
f_1(\bm k) = - e \tau_p \, \frac{d f_0}{d \varepsilon} \, (\bm
v_\| \cdot \bm F) \:,
\]
$e$ is the electron charge, and $\tau_p$ is the momentum
relaxation time.

Substituting the spin vectors of the Dresselhaus eigen-states
$\bm{s}_{\pm}$ in the form Eq.~(\ref{s_dr}) into
Eq.~(\ref{spin_gen_f}), one derives the angular dependence of the
average spin of the transmitted carriers,
\begin{equation}\label{spin}
s_x = \frac{v_{d,x}}{2 \,v_d} P_s \:, \;\; s_y = -
\frac{v_{d,y}}{2\, v_d} P_s \:,
\end{equation}
where $\bm{v}_{d}=(e \tau_p / m^*) \bm{F}$ is the in-plane drift
velocity of the incident electrons, and $P_s$ is the spin
polarization of the transmitted particles,
\begin{equation}\label{p_s}
P_s=\frac{v_d \,m^*}{2 \, \dot{N}} \sum_{\bm k_\|, \, k_z>0}
\frac{d f_0}{d \varepsilon}  \left[ \left| t_{+}(\varepsilon_z,
k_{\|}) \right|^2 - \left| t_{-}(\varepsilon_z, k_{\|})
\right|^2\right] v_z v_{\parallel} \:.
\end{equation}
The direction of the electron spin~(\ref{spin}) is determined by
the symmetry of the Dresselhaus term. In particular, the spin
$\bm{s}$ is parallel (or antiparallel) to the electron drift
velocity $\bm{v}_d$, if $\bm{v}_d$ is directed along the crystal
cubic axis $[100]$ or $[010]$; and $\bm{s}$ is perpendicular to
$\bm{v}_d$, if the latter is directed along the the axis $[1
\bar{1}0]$ or $[1 10]$.

In the limit of thick barriers the structure transparency can be
approximated by Dirac $\delta$-functions. Then substituting the
transparency $|t_{\pm}|^2$ in the form Eq.~(\ref{t}) and assuming
spin-orbit interaction to be small, one derives the following
expression for the spin polarization of the transmitted carriers
\begin{equation}\label{p_s_analyt}
P_s = \frac{v_d \,m^*}{\hbar} \left( \alpha / \zeta - \beta
\right)\:,
\end{equation}
where $\zeta=\int_{E_0}^{\infty} f_0(\varepsilon)d\varepsilon /
f_0(E_0)$ is an energy equal to $E_F-E_0$ for 3D Fermi and $k_B T$
for 3D Boltzmann gas, respectively.

\begin{figure}[b]
\vspace{0.5cm}
\includegraphics[width=0.45\textwidth]{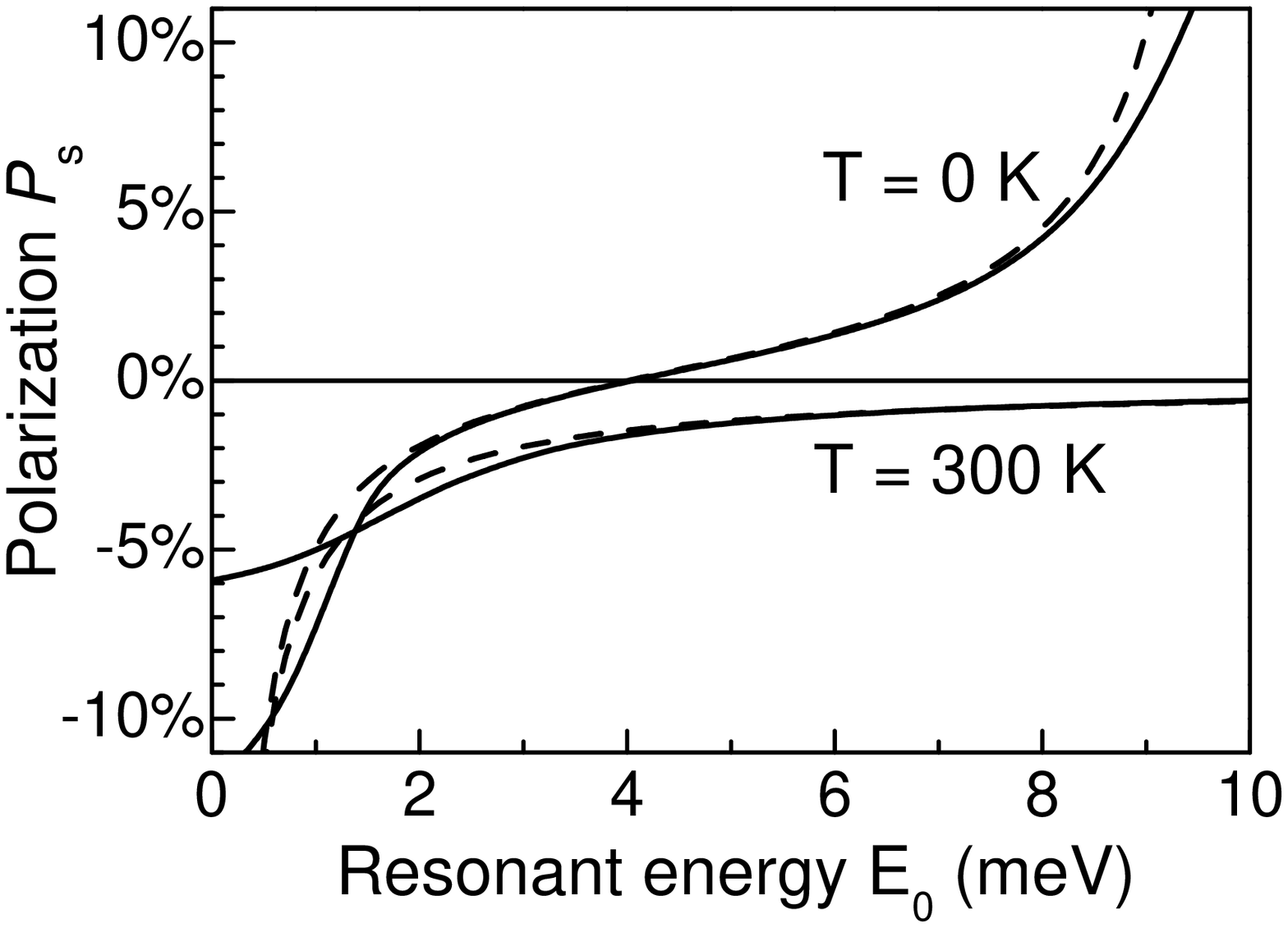}\\ \caption{The spin
polarization $P_s$ as a function of the resonant level position
$E_0$ for degenerate electron gas with $E_F=10$~meV and
non-degenerate gas of the same carrier concentration at $T=300$~K.
Solid curves correspond to the numerical calculation, dashed
curves are plotted following Eq.(\ref{p_s_analyt}). The parameters
of the double-barrier structure are presented in the caption to
Fig.\ref{Fig2}, and $v_d=2.5\cdot10^6$~cm/s that corresponds to
$v_F/10$.}\label{Fig3}
\end{figure}
The dependence of the spin polarization $P_s$ on the resonant
level position $E_0$ is presented in Fig.\ref{Fig3} for the
incident electrons forming degenerate gas with the Fermi energy
$E_F=10$~meV and the gas of the same carrier concentration at
temperature $T=300$~K. Solid curves correspond to the exact
solution of Eq.(\ref{p_s}), with the transmission coefficients
$t_{\pm}(\varepsilon_z,k_{\parallel})$ being calculated
numerically in the framework of the transfer matrices. The
variation of $E_0$ from 0 to 10~meV was achieved modifying the
quantum well width from $a\approx31$\AA \mbox{} to $a\approx
29.4$\AA. Dashed curves are plotted following
Eq.(\ref{p_s_analyt}) for Fermi and Boltzmann statistics. One can
see that in the wide range of the resonant level positions the
simple analytical theory demonstrates a good agreement with the
numerical calculations. For the reasonable set of parameters given
in the figure caption it is possible to achieve spin polarizations
of several percents. The sign of the polarization $P_s$ is
governed by interplay between the spin splitting, $\alpha
k_{\parallel}$, and the difference of the widths of the spin
sublevels, $\beta k_{\parallel}$, since the carrier occupation of
the lower sublevel ``$-$'' is larger than that of the higher
sublevel ``$+$'' while the tunneling transparency of the sublevel
``$-$'' is smaller than that of ``$+$'' (see Eq.(\ref{t})). It
clarifies why the terms proportional to $\alpha$ and $\beta$
contribute to Eq.(\ref{p_s_analyt}) with opposite signs. At low
temperature both terms are comparable, and $P_s$ changes the sign
with increasing of $E_0$. At $E_0 \ll E_F$ the effect is mainly
determined by the difference of the spin sublevel transparencies.
The role of the energy spacing between the sublevels is
negligible, since the carrier populations at the energies $E_{+}$
and $E_{-}$ almost coincide. With increasing of $E_0$ the
``step-like'' Fermi distribution leads to different occupations of
the spin sublevels, and the polarization $P_s$ changes the sign.
At higher temperatures the carrier distribution becomes to be
smooth, and the effect is related mainly to the difference of the
tunneling transparencies of the spin sublevels.
\section{Tunneling spin galvanic effect}\label{detection}
Generation of an electric current by spin-polarized carriers
represents the effect inverse to spin injection. Now we assume the
electron gas on the left side of the double-barrier structure to
be spin-polarized. Electrons with various wave vectors tunnel
through the RTD. However, due to the splitting of the resonant
level the flux of the spin-polarized carriers with the certain
in-plane wave vector $\bm{k}_{\parallel}$ is larger than the flux
of the particles with the opposite in-plane wave vector,
$-\bm{k}_{\parallel}$. This asymmetry results in the in-plane flow
of the transmitted electrons near the barrier, i.e. in the
interface electric current. The direction of this interface
current is determined by the spin orientation of the electrons and
symmetry properties of the barrier, in particular the current
reverses its direction if the spin orientation changes the
sign.\cite{Hall,Tarasenko}

The theory of such ``tunneling spin-galvanic effect'' is developed
by using the spin density matrix technique. The interface current
of spin-polarized electrons transmitted through the tunneling
structure is given by~\cite{Tarasenko}
\begin{equation}\label{j_gen}
\bm{j}_{\parallel}=e \sum_{\bm k_\|, \, k_z>0} \tau_p \,
\bm{v}_{\parallel} \, v_z \mbox{Tr} \left[ {\cal T} \rho_l {\cal
T}^{\dag} \right]  \:,
\end{equation}
where $\tau_p$ is the momentum relaxation time, $\rho_l$ is the
electron density matrix on the left side of the structure, and
${\cal T}$ is the spin matrix of the tunneling transmission that
links the incident spinor wave function $\psi_l$ to the
transmitted spinor wave function $\psi_r$, $\psi_r = {\cal T}
\psi_l$. We assume the carriers on the left side of the structure
to form 3D spin-oriented electron gas, and electron distributions
in both spin subband to be thermalized. For the case of small
degree of spin polarization, the density matrix has the form
\begin{equation}\label{rhol}
\rho_l \approx f_0 \hat{I} -  \frac{d f_0}{d \varepsilon} \frac{2
p_s}{\langle 1/ \varepsilon \rangle} \,(\bm{n}_s \cdot
\hat{\bm{\sigma}}) \:,
\end{equation}
where $f_0$ is the equilibrium distribution function of
non-polarized carriers, $\bm{n}_s$ is the unit vector directed
along the spin orientation, $p_s$ is the degree of the
polarization, and $\langle 1/ \varepsilon \rangle$ is the average
value of the reciprocal kinetic energy of the carriers equal to
$3/E_F$ for 3D degenerate electron gas with the Fermi energy
$E_F$, and $2/ k_B T$ and 3D non-degenerate gas at the temperature
$T$. The spin matrix of the electron transmission through the
structure is given by
\begin{equation}\label{T_matr}
{\cal T} = t_{+}(\varepsilon_z,k_{\|}) \, \chi_{+} \chi^{\dag}_{+}
+ t_{-}(\varepsilon_z,k_{\|}) \,\chi_{-} \chi^{\dag}_{-} \:.
\end{equation}
Substituting the density matrix~(\ref{rhol}) and the transmission
matrix~(\ref{T_matr}) into Eq.(\ref{j_gen}) and taking into
account the definition of the vectors $\bm{s}_{\pm}$~(\ref{s_dr}),
one obtains
\[
\bm{j}_{\parallel} = -\frac{4 e \, \tau_p \,p_s}{\langle 1/
\varepsilon \rangle} \sum_{\bm k_\|, \, k_z>0} \frac{d f_0}{d
\varepsilon} \left[ \left| t_{+}(\varepsilon_z, k_{\|}) \right|^2
\bm{n}_s \cdot \bm s_{+}(\bm k_{\|}) \right.
\]
\vspace{-5mm}\begin{equation} +\, \left. \left|
t_{-}(\varepsilon_z, k_{\|}) \right|^2 \bm{n}_s \cdot \bm
s_{-}(\bm k_{\|}) \right] \bm{v}_{\|} \,v_z \:.
\end{equation}

Taking into account the explicit form of the spin vectors of the
Dresselhaus eigen-states~(\ref{s_dr}), the components of the
tunneling spin-galvanic current have the form
\begin{equation}\label{jtsge}
j_{\parallel, x} = - j_{\parallel} \, n_{s,x} \: , \;\;
j_{\parallel, y} = j_{\parallel} \, n_{s,y} \:,
\end{equation}
where
\begin{equation}\label{j_int}
j_{\parallel} = - \frac{e\tau_p p_s}{\langle 1/ \varepsilon
\rangle} \sum_{\bm k_\|, k_z>0} \frac{d f_0}{d \varepsilon} \left[
\left| t_{+}(\varepsilon_z, k_{\|}) \right|^2 - \left|
t_{-}(\varepsilon_z, k_{\|}) \right|^2\right] v_z v_{\parallel}
\:.
\end{equation}
The direction of the tunneling spin-galvanic current is determined
by the spin orientation of the electrons with respect to the
crystallographic axes.

In the limit of thick barriers the structure transparency can be
obtained substituting $\delta$-functions~(\ref{t}) for the
structure transparency. Then assuming spin-orbit interaction to be
small, the interface current is derived to be
\begin{equation}\label{j_simp}
j_{\parallel} = - \frac{e\,\tau_p \,p_s}{\langle 1/\varepsilon
\rangle} \frac{\,f(E_0)\,m^*}{\pi \hbar^4} \left(\alpha -
\zeta\beta \right)\Gamma_0 \:.
\end{equation}

\begin{figure}[t]
\vspace{0.3cm}
\includegraphics[width=0.45\textwidth]{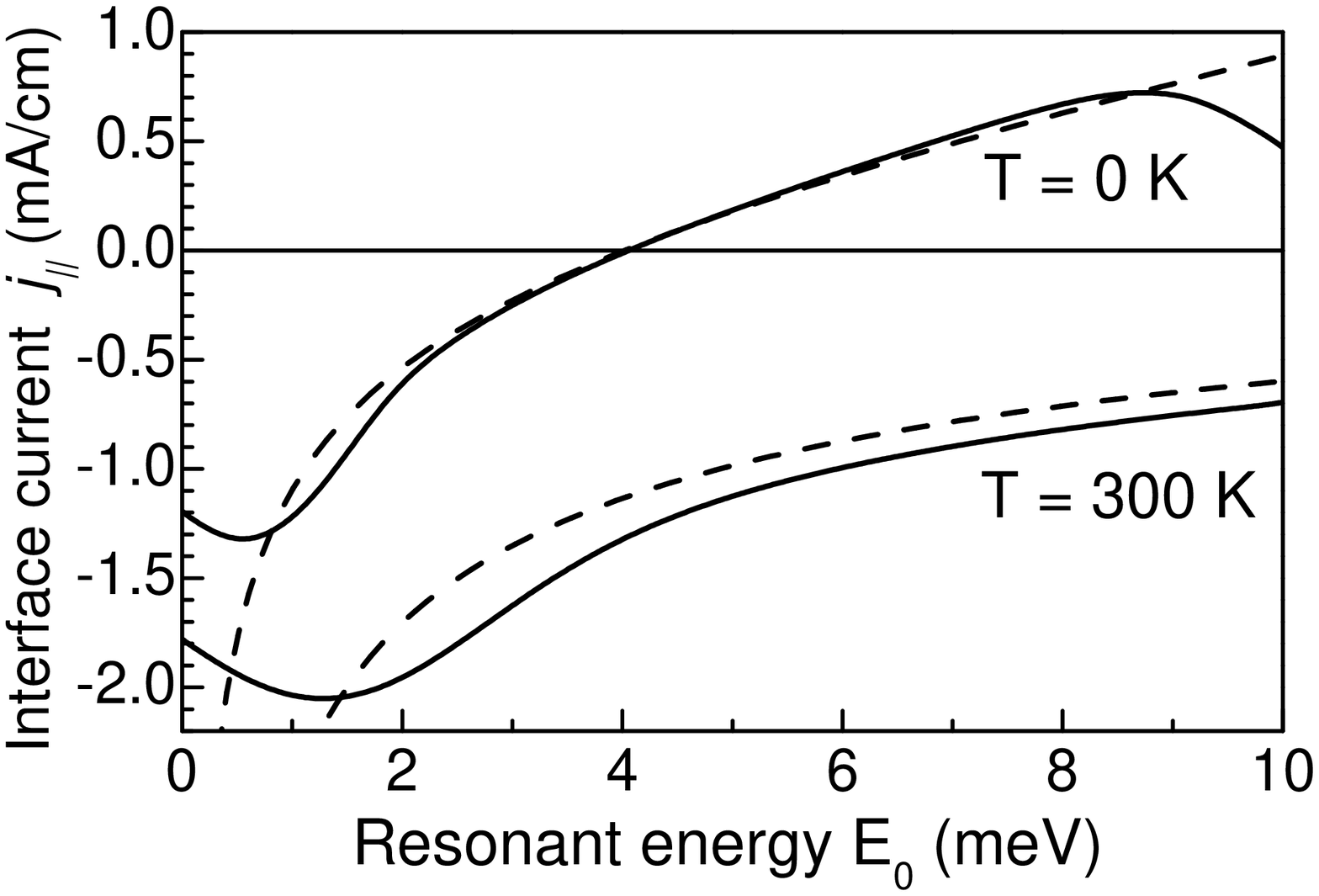}\\
\caption{The interface current $j_\parallel$ as a function of the
resonant level position $E_0$ for degenerate electron gas with
$E_F=10$~meV and non-degenerate gas of the same carrier
concentration at $T=300$~K. Solid curves correspond to the
numerical calculation, dashed curves are plotted following
Eq.(\ref{j_simp}). The parameters of the double-barrier structure
are presented in the caption to Fig.\ref{Fig2}, $\tau_p = 1$~ps,
and $p_s=0.1$.}\label{Fig4}
\end{figure}

Fig.\ref{Fig4} presents the dependence of the interface current
$j_{\parallel}$ on the energy position of the resonant level $E_0$
for the incident electrons forming spin-polarized degenerate gas
with the Fermi energy $E_F=10$~meV and the gas of the same carrier
concentration at temperature $T=300$~K. Solid curves correspond to
the exact solution of Eq.(\ref{j_int}), with the transmission
coefficients $t_{\pm}(\varepsilon_z,k_{\parallel})$ being
calculated numerically. Dashed curves are plotted following
Eq.(\ref{j_simp}) for Fermi and Boltzmann statistics. For the
AlGaSb-based RTD structure considered as an example the tunneling
spin-galvanic current is of order of mA/cm. Estimation shows that
it is enhanced by an order of magnitude with respect to the
interface current generated under tunneling through a single
AlGaSb-based barrier provided the equal electron tunneling flux,
$\dot{N}\sim 10^{22}$~1/(cm$^{2}\,$s).~\cite{Tarasenko} Similarly
to the spin polarization~(Fig.\ref{Fig3}), the sign of the
interface current~(\ref{j_simp}) is governed by interplay between
the contributions responsible for the spin splitting and the
difference of the spin sublevel transparencies.

In conclusion, the theory of spin-dependent electron tunneling has
been developed for symmetrical double-barrier structures based on
zinc-blende-lattice semiconductor compounds. The Dresselhaus
spin-orbit interaction couples spin states and space motion of
conduction electrons that leads to spin splitting of the resonant
level depending on the in-plane electron wave vector. The effect
of the spin-dependent tunneling could be employed for creating
spin injectors and detectors based on nonmagnetic tunneling
structures.

This work was supported by the RFBR, the INTAS, programs of the
RAS, Russian Scientific Schools, and Foundation "Dynasty" - ICFPM.


\begin{thebibliography}{16}
\bibitem{non_parabolicity} The effective mass and the Dresselhaus constant
can be estimated from GaSb band parameters at the conduction band
bottom, $m^* = 0.041 m_0$ and $\gamma =
187$~eV$\cdot$\AA$^3$~[\onlinecite{Ivchenko_Pikus}], taking into
account non-parabolicity.\cite{Pikus_Titkov}
\end{thebibliography}
\end{document}